\documentclass[%
 reprint,
amsmath,amssymb,
aps,
floatfix,
]{revtex4-2}

\usepackage{graphicx}
\usepackage{dcolumn}
\usepackage{bm}
\usepackage[colorlinks,urlcolor=blue,citecolor=blue,linkcolor=blue]{hyperref}
\usepackage{svg}
\usepackage{braket}

\newcommand{\lrangle}[1]{\langle #1 \rangle}
\newcommand{\omegaratio}{\Omega_{2}^{2}/\Omega_{1}^{2}}
\newcommand{\fprime}{F^{\prime}}

\begin{document}
\preprint{APS/123-QED}

\title{Doppler Shift Mitigation in a Chip-Scale Atomic Beam Clock}

\author{Alexander Staron,$^{1,2,\dagger}$ Gabriela Martinez,$^{1,2,\ast}$ Nicholas Nardelli,$^{1,2}$ Travis Autry,$^{3}$ John Kitching,$^{1}$ and William McGehee$^{1}$}
\affiliation{$^{1}$Time and Frequency Division, National Institute of Standards and Technology, Boulder, CO, USA}
\affiliation{$^{2}$Department of Physics, University of Colorado Boulder, Boulder, CO, USA}
\altaffiliation{Present Address: Mesa Quantum, Boulder, CO, USA}
\affiliation{$^{3}$HRL Laboratories, Malibu, CA, USA}
\email{Contact author: alexander.staron@colorado.edu}

\date{\today}

\begin{abstract}
Chip-scale microwave atomic systems based on thermal atomic beams offer a promising approach to realize low-power and low-drift clocks for timing holdover applications. Miniature beam clocks are expected to suppress many of the shifts that commonly limit existing chip-scale atomic clocks based on coherent population trapping, including collisional shifts and some light shifts. However, the beam geometry can amplify some challenges such as Doppler shifts, which generate a strong sensitivity to laser frequency variation. Using a cm-scale $^{87}$Rb atom beam clock, we identify a surprisingly strong competition between Doppler shifts and resonant light shifts arising from asymmetric decay in the clock spectroscopy $\Lambda$-system. Leveraging this competition between Doppler and resonant light shifts, we demonstrate clock operation at specific, convenient experimental parameters consistent with zero sensitivity to laser frequency variation and white-noise-limited clock frequency averaging for 1000 s of integration. 
\end{abstract}

\maketitle

\section{\label{intro}Introduction}
Chip-scale atomic clocks (CSACs) based on coherent population trapping (CPT) resonances in atomic vapors are low-power ($\sim 0.1 \ {\rm W}$) devices capable of maintaining $\mu$s-level time accuracy without steering over approximately one day~\cite{vanier_atomic_2005, kitching_chip-scale_2018}. The instability of CSACs after one day is usually limited by a combination of light shifts and aging of the buffer gas environment within the vapor cell. To achieve microsecond-level timing over a week or longer, atomic clocks with significantly higher power ($>~5 \ {\rm W}$) and volume than existing CPT-based approaches are used~\cite{marlow_review_2021}. Efforts to significantly reduce the drift of CSACs have been explored using vapor cells with lower gas permeability to reduce cell aging~\cite{abdel_hafiz_light-shift_2022,carle_reduction_2023} and pulsed interrogation to mitigate light shifts~\cite{carle_pulsed-cpt_2023}, but fully miniaturized clocks that take advantage of these improvements have yet to be realized. 

Recently, methods to miniaturize atomic beams~\cite{li_cascaded_2019,li_robust_2020} have enabled the demonstration of a chip-scale atomic beam clock (CSABC)~\cite{martinez_chip-scale_2023} based on Ramsey-CPT spectroscopy. Motivated by the rich history of thermal beams in atomic timekeeping~\cite{ramsey_molecular_1950, essen_atomic_1955, vanier_quantum_1988, shirley_accuracy_2001}, the CSABC approach has demonstrated similar short-term stability to existing CSACs and is expected to have a lower sensitivity to changes in the vacuum environment and reduced light shifts relative to continuously-interrogated, vapor cell-based CPT clocks. This makes the CSABC a promising approach for low-drift timing while maintaining the low power consumption of a traditional CSAC. Miniature atomic beams have also been used to explore stimulated laser cooling~\cite{li_stimulated_2023} and the generation of non-classical light~\cite{larsen_chip-scale_2025} at the chip scale. 

The stability and systematics of Ramsey-CPT atomic beam clocks have been studied in detail using laboratory-scale experiments ~\cite{thomas_observation_1982,hemmer_precision_1986,hemmer_ac_1989,hemmer_semiconductor_1993}. Phase shifts arising from incomplete formation of the CPT dark state, called resonant light shifts, are a leading source of clock instability and can depend strongly on the common-mode laser detuning. These resonant shifts typically arise from a population imbalance between the ground states of the $\Lambda$-system and can be exponentially suppressed by strongly saturating the CPT transition~\cite{hemmer_ac_1989, shahriar_darkstate_1997, blanshan_light_2015}. Light shifts also arise from the off-resonant laser frequency components. The magnitude of this light shift is determined by the relative strength of the off-resonant transitions and is largely unaffected by laser detuning~\cite{zhu_theoretical_2000, levi_light_2000}. These shifts can be minimized for specific ratios of CPT field intensities and can be independent of total laser intensity using Ramsey spectroscopy~\cite{vanier_quantum_1988, zhu_theoretical_2000, pollock_ac_2018}. 

First-order Doppler shifts also arise in the CSABC approach and depend on the common-mode laser detuning. The Doppler shift is analogous to the ``end-to-end cavity phase shift” observed in cavity-based microwave clocks~\cite{shirley_accuracy_2001,jefferts_2002} as the microwave phase depends on the position of the atoms in the CPT laser fields. For the simplest Ramsey CPT approach with light incident from one direction, the Doppler sensitivity is set by the ratio of the CPT optical and microwave wavelengths at $\approx 18.1 \ {\rm Hz/MHz}$ for $^{87}$Rb. To reach useful levels of clock performance, the Doppler shift requires  $\sim 1 \ {\rm kHz}$ laser stability, which is likely infeasible with lower-power laser systems such as those used in the CSAC~\cite{serkland_vcsels_2007}. Interrogation using counter-propagating CPT fields can reduce the Doppler sensitivity at the cost of increased experimental complexity~\cite{jau_push-pull_2004}. 


\begin{figure*}
\includegraphics[width=\textwidth]{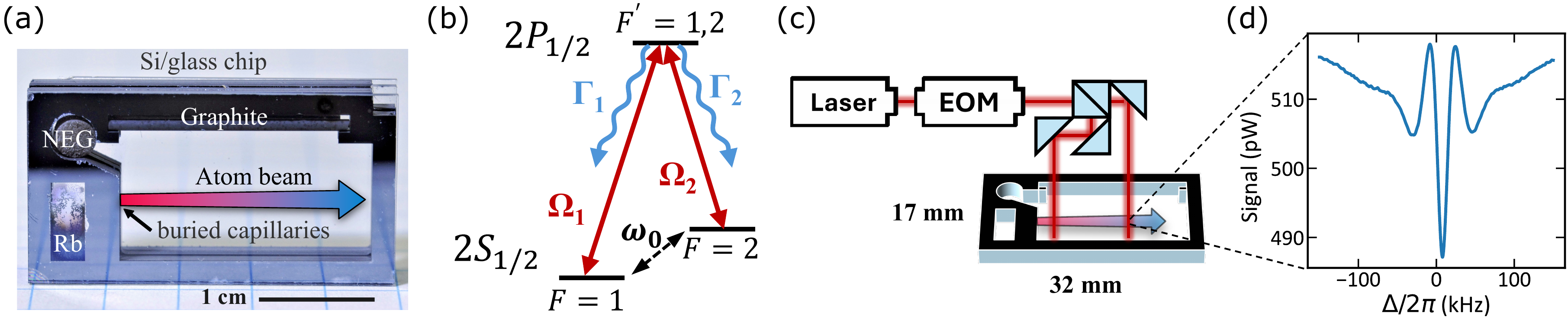}
\caption{Ramsey CPT spectroscopy using a chip-scale atomic beam. (a) Image of the atomic beam device showing internal components and the location of the atomic beam (cartoon arrow). (b) Energy level diagram of the $^{87}$Rb, D1 lines indicating the relevant atomic levels, CPT Rabi rates (red), and decay rates (blue). (c) Schematic of light and optics shows how the atomic beams are probed. (d) Typical Ramsey fringe measured relative to the Raman detuning $\Delta$. }
\label{device_schematic}
\end{figure*}

Here, we perform Ramsey-CPT spectroscopy on the ground-state hyperfine transition in $^{87}$Rb using a chip-scale atomic beam device to explore the stability limits of the CSABC approach. We study the sensitivity of the clock frequency to variations in laser detuning, and we find that a competition between Doppler shifts and light shifts exists which can enhance or null the clock's total laser detuning sensitivity. This sensitivity varies with the CPT field intensities, depends on the atomic levels probed, and tunes to zero using a convenient set of experimental parameters. The light shifts are shown to be consistent with resonant phase shifts arising from asymmetric decay in the CPT process~\cite{hemmer_ac_1989}, and simple modeling that accounts for the spread of transverse laser detunings can largely reproduce the observed behavior. The reduced laser detuning sensitivity is then utilized to demonstrate improved long-term clock stability.

\section{Chip-scale beam clock}
The chip-scale beam device used in this experiment (see Fig.~\ref{device_schematic}(a)) produces Rb atomic beams which propagate in a passively pumped vacuum environment. The device is fabricated from a five-layer stack of anodically-bonded Si and glass to form a hermetic cell with a total volume of $\approx 3.5 \ {\rm mL}$, based on the design presented in Ref.~\cite{martinez_chip-scale_2023} and producing similar atomic beam flux and vacuum conditions. The device contains an internal ``source" cavity containing Rb metal connected to a lithographically etched microcapillary array~\cite{li_cascaded_2019, li_robust_2020} for producing atomic beams in a larger ``drift" cavity used for atom beam spectroscopy. 

The microcapillary array contains ten 100 $\mu$m-width square channels, each with an aspect ratio of 30. When heated, Rb vapor feeds the microcapillary array to generate a set of atomic beams with a total flux of $10^{12} \ {\rm s}^{-1}$ at $\approx 373 \ {\rm K}$. The atomic beams cover a 1.45 mm by 100 $\mu$m area at the output of the capillaries, and the beams expand with a divergence of $\approx 30 \ \rm{mrad}$, set by the aspect ratio of the microcapillaries. This produces a one-photon fluorescence signal with full width at half maximum of $\approx 40 \ \rm{MHz}$ in the detection region located $\approx 11 \ \rm{mm}$ from the channel exit~\cite{martinez_chip-scale_2023}. Non-evaporable getters (NEGs) and graphite rods are used to maintain the vacuum environment in the drift cavity at a pressure  $\textless \ 1 \ {\rm Pa}$~\cite{martinez_chip-scale_2023}. The atom beam flux is consistent with molecular flow models~\cite{beijerinck_velocity_1975}, indicating that collisional losses are not significant at this pressure.


Ramsey-CPT spectroscopy of the hyperfine ground state levels of $^{87}$Rb with clock frequency $\omega_0 \approx 2\pi \times 6.835 \ {\rm GHz}$ is performed using the $D_{1}$ transitions ($\lambda \approx 795 \ {\rm nm}$, natural linewidth $\Gamma \approx 2 \pi \times 5.75 \ {\rm MHz}$) addressing excited states $F^{\prime} = 1,2$ as shown in Fig.~\ref{device_schematic}(b). The CPT light is derived from a distributed feedback laser stabilized on the $F = 2\rightarrow F^{\prime}$ transition with Rabi rate $\Omega_2$. The light is modulated using a fiber electro-optical phase modulator (EOM) driven near $\omega_0$ to produce light resonant with the $F = 1$ to $F^{\prime}$ transitions with Rabi rate $\Omega_1$. We measure the power ratio of the resonant laser components, $\omegaratio$, using a Fabry-Pérot optical spectrum analyzer. This ratio is near one for most CPT experiments which use laser modulation at $\omega_0/2$, but full-hyperfine modulation is used so that the ratio $\Omega_2^2/\Omega_1^2$ and its impact on light shifts can be measured. The common-mode laser detuning $\delta_L$ is controlled by offsetting the laser lock using an acousto-optic modulator. 

Two parallel laser beams with a nominal one centimeter separation probe the atomic beams. The beams are derived from a single laser beam split into two paths such that the propagation distance to the two Ramsey zones is approximately equal. The path symmetry minimizes the microwave phase difference on the CPT light between the zones and is realized using a non-polarizing beam-splitter and three right-angle mirrors as shown in Fig.~\ref{device_schematic}(c).  The beam-splitting optics are mounted to a temperature-stabilized, miniature optical bench made of Si to reduce phase instability from thermal effects. The circularly-polarized beams have $1/e^{2}$ radii $\approx 500 \ (1400) \ \mu{\rm m}$ along (normal to) the atomic beam axis, chosen to uniformly address atoms in the first zone and minimizes Rabi pulling. 

The optical power is split between the two Ramsey zones with $\approx 75 \ \%$ of the light sent to the first zone. This splitting provides strong pumping in the first Ramsey zone and weak read-out in the second zone, which is commonly employed in Ramsey-CPT spectroscopy~\cite{blanshan_light_2015, micalizio_Raman-Ramsey_2019} and is empirically found to maximize the clock signal-to-noise ratio. We define the resonant optical power in this first zone as $P_{\rm res}$. A magnetic field of $\approx 0.36  \ {\rm mT}$ is applied along the $k$-vector of the light to separate the magnetically-sensitive ($m_{F} = \pm 1$) $\Lambda$-systems from the clock state ($m_{F} = 0$). The Ramsey beams propagate normal to the atomic beam axis.  In order to minimize back-reflected light on the probed atoms, the beams are incident at $40^{\circ}$ relative to cell's surface normal. 

We measure fluorescence from the second Ramsey zone using a high-gain Si photodiode (1~mm active area) to produce the 15~kHz full width at half maximum Ramsey fringe shown in Fig.~\ref{device_schematic}(d). The fringe width is set by the average Ramsey dark time $T_R \approx 30 \ \mu {\rm s}$. A 1:1 imaging system collects the fluorescence with $> 10 \ \%$ efficiency and rejects scattered light, enabling signal contrast in the $5 - 10 \ \%$ range depending on the cell temperature, $F^{\prime}$, and $P_{\rm res}$. A clock is formed by stabilizing the EOM modulation frequency $\omega$ onto the $m_F = 0$ Ramsey fringe using a maser as a frequency reference. The clock frequency is given by $\Delta = \omega - \omega_{0}$, and the clock's fractional frequency stability is roughly $1\times10^{-9}$ at 1 s of integration for $F^{\prime} = 2$ at $373 \ {\rm K} \ (100 \ {\rm C})$. The short-term stability improves with cell temperature via increased atomic flux.

\section{Laser Detuning Sensitivity}

\begin{figure}[h]
\centering
\includegraphics[width=\linewidth]{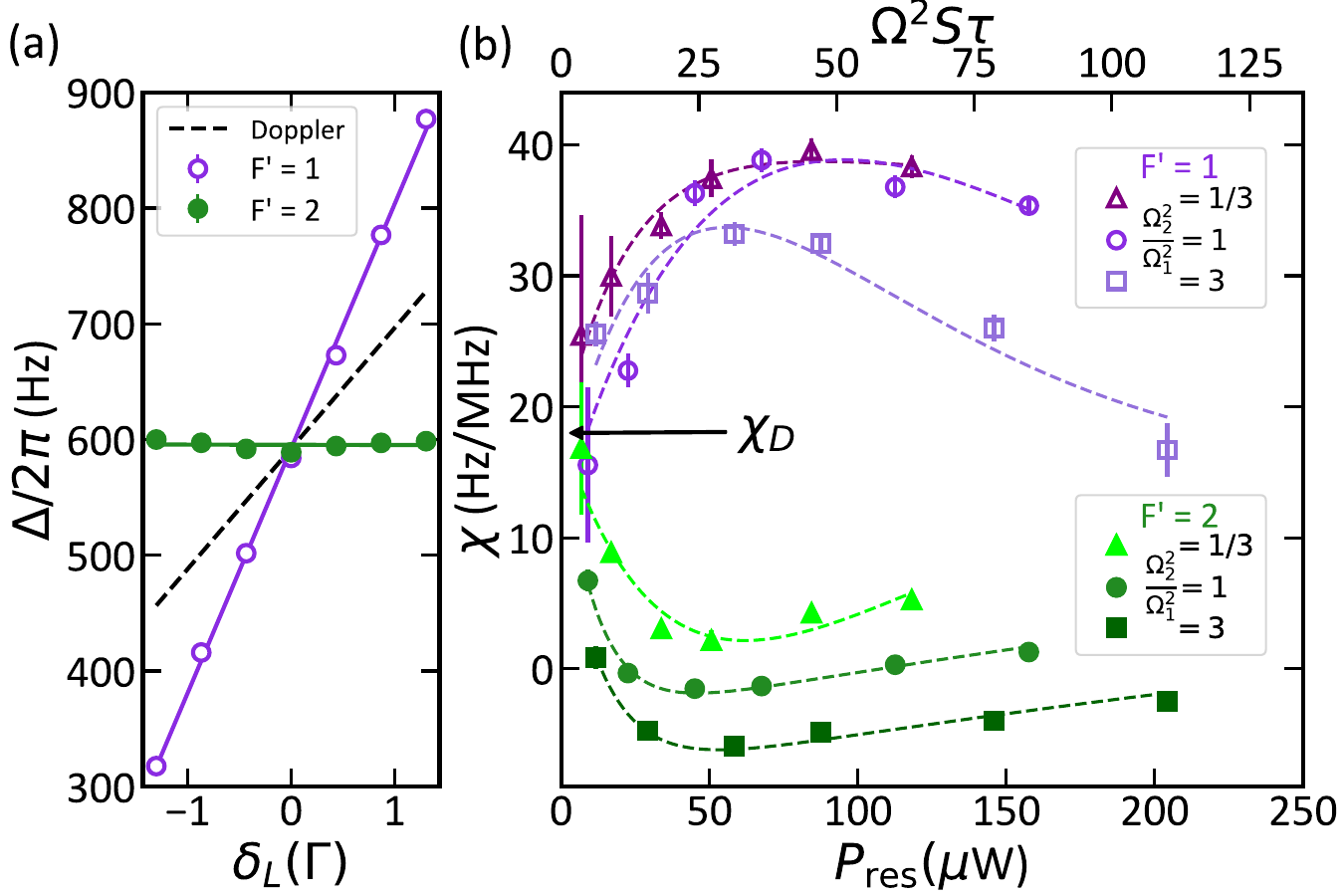}
\caption{Clock sensitivity to laser detuning. (a) Clock frequency is measured vs. laser detuning for $\fprime = 1$ (purple) and $\fprime = 2$ (green) using $P_{\rm res} \approx 113 \ \mu{\rm W}$ and $\omegaratio \approx 1$. Solid lines are linear least squares fits used to determine $\chi$. (b) Measured $\chi$ vs. laser power for $\fprime = $ 1~(purple), $F^{\prime} = 2$~(green) and $\Omega_{2}^{2}/\Omega_{1}^{2} \approx $ 0.33~(triangles), 1~(circles), and 3~(squares). Dashed lines are guides to the eye. The Raman damping parameter $\Omega^{2}S\tau$ is calculated for $F' = 2$, $\tau =  2 \ \mu {\rm s}$, and $\delta = 0$ (values are three times lower for $F' = 1$). Laser detuning sensitivity adds (subtracts) from $\chi_D$ for $F^{\prime}$ = 1(2) due to light shifts, and the sensitivity is shown to cross zero for $F^{\prime} = 2$ at certain values of $P_{\rm res}$. Error bars indicate the standard error of the mean; some are smaller than their marker.}
\label{fig:power}
\end{figure}
We investigate the sensitivity of the clock frequency to changes in common-mode laser detuning over a range of experimental conditions using the $\fprime = 1,2$ excited states. This sensitivity is expected to be the largest source of instability in CSABCs, and we define the detuning sensitivity $\chi = \partial\Delta / \partial\delta_L|_{\delta_L = 0}$. We measure $\chi$ using linear fitting of the clock frequency over a range of $\delta_L$ covering $\pm \Gamma$ as shown in Fig.~\ref{fig:power}(a). For this plot, the 2nd-order Zeeman shift is measured and subtracted from $\omega$, leaving a residual $\approx 600 \ {\rm Hz}$ clock offset at zero detuning. This offset could be attributed to a combination of microwave phase shifts from relative beam misalignment or light shifts from off-resonant components in the CPT fields. The ratio of the CPT Rabi rates, $\omegaratio$, is set by the modulation index of the EOM.

The observed detuning sensitivity differs significantly from a straightforward, constant Doppler shift at $\chi_{\rm D} \approx 18.1 \ {\rm Hz/MHz}$, as indicated by the black arrow in Fig.~\ref{fig:power}(b). The sensitivity approaches $\chi_{\rm D}$ at low $P_{\rm res}$ before rapidly changing with drive power. The sensitivity differs greatly between the $\fprime$ states, with values significantly below $\chi_D$ (and near zero) for $\fprime = 2$ and significantly above $\chi_D $ for $\fprime = 1$ at intermediate values of $P_{\rm res}$. The maximal deviation from $\chi_{D}$ occurs near $\Omega^2S\tau \approx 20$ for both excited states. At the highest values of $P_{\rm res}$, $\chi$ exhibits slow decay toward $\chi_{D}$. The detuning sensitivity also varies as the ratio of Rabi rates is adjusted, tending to decrease with higher ratios of $\omegaratio$. We attribute the observed behavior to resonant light shifts, whose laser frequency dependence $\chi_{\rm res}$ is discussed below and adds or subtracts from the Doppler shift as $\chi = \chi_{D} + \chi_{\rm res}$. 

The zero crossings for $\chi$ using $F^{\prime} = 2$ provide convenient locations for clock operation with greatly reduced sensitivity to laser frequency drift. The locations of these crossings varies with both optical power and $\omegaratio$, making the crossings flexible with respect to experimental conditions. For example, it is often desirable to operate chip-scale clocks using modulation at $\omega_0/2$ such that $\omegaratio \approx 1$, and zero-crossings are observed in that configuration at reasonable optical power levels. 

\section{Resonant Light Shifts}

To explain the $\chi$ behavior observed in Fig.~\ref{fig:power}, we rely on the understanding of light shifts developed for Ramsey-CPT spectroscopy in Ref.~\cite{hemmer_ac_1989}. Light shifts in CPT clocks generally include conventional ac Stark shifts, which scale linearly with optical power, but larger phase shifts, called resonant or coherent light shifts, can form as the atoms are pumped into the CPT dark state. These shifts have been studied primarily in the context of initial population differences between the ground states~\cite{hemmer_ac_1989,blanshan_light_2015}, where the sign of the shift depends on the relative occupation of the two ground state levels before the CPT pumping begins $\Delta\rho_{0} = \rho^{0}_{11} - \rho^{0}_{22}$, where $\rho^0_{11} (\rho^0_{22})$ are the initial $F = 1(2)$ occupations. This initial population imbalance generates a phase shift $\phi_{\rm res}$ which is initially maximal and exponentially decreases as the atoms are pumped into phase with the driving fields during the CPT process.

\begin{figure}[ht]
\centering
\includegraphics[width=\linewidth]{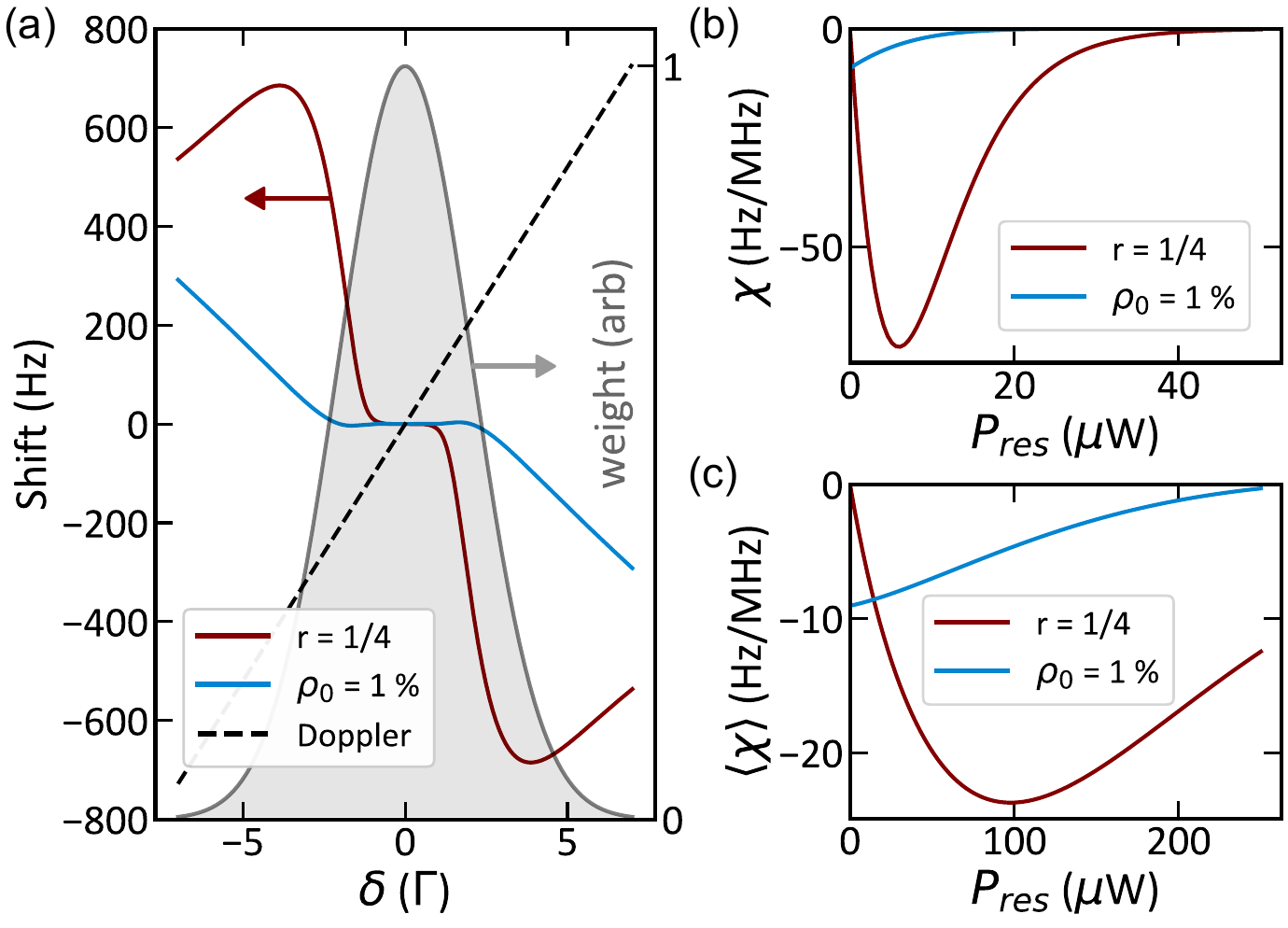}
\caption{Resonant light shifts and laser detuning sensitivity. (a) The resonant light shift is plotted relative to common-mode laser detuning in the limiting cases of $r = 1/4$, $\Delta\rho_0 = 0 $ (red) and $r = 0$, $\Delta\rho_0 = 1 \ \%$ (blue) for $P_{\rm res} = 100 \ \mu{\rm W}$. The Doppler shift (dashed line) is $\approx 18.1 \ {\rm Hz/MHz}$, and the relative line pulling weight for each detuning class in the measured geometry is indicated by the shaded area. The clock sensitivity to laser detuning using the experimental parameters is plotted for these cases considering the behavior near $\delta = 0$ (b) and the sensitivity averaged over the line pulling weights (c) using Eq.~(\ref{eq:<chi>}).}
\label{fig:theory}
\end{figure}

An analytical approximation for the pumping-dependent resonant phase shift arising from an initial population imbalance can be derived in the limit of equal Rabi rates, weak pumping, and equal excited state decay rates~\cite{hemmer_ac_1989}  as 
\begin{equation}\label{eq:res_rho}
\tan(\phi_{\rm res}) =  -\Delta\rho_{0} \ \frac{ \sin(\Omega^2\!S\tau\delta/\Gamma ) }{e^{\Omega^2\!S \tau}-1},
\end{equation}
where $\Omega^2\!S =  \frac{1}{2}(\Omega_1^2 + \Omega_2^2)\frac{\Gamma}{\Gamma^{2} + 4\delta^{2}}$ is the Raman damping rate in the first Ramsey zone, $\delta$ is the common-mode detuning from atomic resonance, and $\tau$ is the interaction time in the first Ramsey zone. The weak pumping assumption ($\Omega^2\!S \ll \Gamma$) is valid for many of the atoms probed due to the spread of measured Doppler detunings ($\delta = \delta_L - k v_{\perp}$,  where $v_{\perp}$ is the atomic velocity along the laser k-vector and $k = 2\pi/\lambda$). This shift and its derivative at $\delta = 0$ are plotted in blue in Fig.~\ref{fig:theory}(a) and Fig.~\ref{fig:theory}(b), respectively, for a hypothetical difference in ground state populations $\Delta\rho_0 =  1 \ \%$ and $T_R = 30 \ \mu {\rm s}$. The resonant light shift is dispersive near $\delta = 0$ and the clock sensitivity to common-mode laser detuning, $\chi$, decays rapidly with driving laser power. The sign of the shift is proportional to $\Delta\rho_0$ and the magnitude of the shift is expected to be near zero in our CSABC as the ground state populations in our thermal atom source are nominally equal. 

Resonant light shifts can also arise from imbalanced decay in the CPT $\Lambda$-system~\cite{hemmer_ac_1989, taichenachev_theory_2003, zanon-willette_ultrahigh-resolution_2011, liu_light-shift_2021, tsygankov_branching_2024} due to the redistribution of population between the ground states arising from the CPT pumping process itself. This imbalanced decay can happen with or without an initial population imbalance, and the effect of this imbalance on resonant light shifts largely explains the strong changes in laser detuning observed in Fig.~\ref{fig:power}. To understand this process, the decay asymmetry is parameterized by the normalized decay rate difference $r = (\Gamma_1 - \Gamma_2)/(\Gamma_1 + \Gamma _2)$ (see Fig.~\ref{device_schematic}(b)), and a simple analytical expression for the resonant phase shift can be found by simplifying Eq. (8a) from Ref.~\cite{hemmer_ac_1989}. In the limit of $\rho_0 = 0$, equal Rabi rates, and weak pumping, the Ramsey phase shift is 
\begin{equation}\label{eq:phi_r}
\tan(\phi_{\rm res}) = -r\frac{2\Gamma}{\delta} \
\frac{\sin^{2}(\frac{1}{2}\Omega^2S\tau\delta/\Gamma)}{e^{\Omega^2S\tau} - 1}.
\end{equation}
This resonant phase shift differs from the $\rho_0$-based shift in Eq.~(\ref{eq:res_rho}) as it is initially zero when the atoms enter the first Ramsey zone and grows during the optical pumping process. The shift and clock sensitivity $\delta = 0$ are shown by the red lines in Fig.~\ref{fig:theory}(a) and Fig.~\ref{fig:theory}(b), respectively, for a hypothetical $r = 1/4$ and $T_R = 30 \ \mu{\rm s}$.  Like the $\rho_0$-based shift, the sign of the $r$-based shift depends on the population imbalance induced by asymmetric decay ($r \ne 0$), and the shift decays exponentially at high Raman damping rates.

We expect $r \ne 0 $ for the CPT $\Lambda$-systems we probe, arising from absence of decay between $m_F = 0$ levels when $F = F^{\prime}$. Under the assumption of equal population in all ground state hyperfine Zeeman sub-levels and pumping using circularly-polarized light, we estimate the values of $r \approx -0.5(0.25)$ for interrogation using $F' = 1 (2)$. These values are only estimates and would likely vary during the CPT process as population redistributes among the Zeeman sub-levels. For the geometry used in these measurements, atoms spend $\tau \approx 2 \ \mu{\rm s} \ (\approx 70/\Gamma)$  in the first Ramsey zone, and atoms resonant with the laser fields likely reach a steady-state population equilibrium. 

The shape and sign of the laser detuning sensitivities presented in Fig.~\ref{fig:power} are consistent with the estimated sign and shape of the $r$-based resonant light shift. Here, we expect $\chi_{\rm res} > 0 $ for $\fprime = 1$ and $\chi_{\rm res} < 0 $ for $\fprime = 2$ due to the difference in the sign of $r$. Fig.~\ref{fig:power} also shows that $\chi$ depends somewhat on $\omegaratio$. This variation differs in sign and magnitude from the weak dependence predicted from the resonant light shift model~\cite{hemmer_ac_1989}, but could arise from modification to $r$ induced by the unequal Rabi rates which we predict would shift $\chi$ in the observed direction. The population dynamics of the CPT $\Lambda$-system have been shown to not only depend on the spontaneous decay channels present, but also on the relative strengths of the driving fields~\cite{jyotsna_coherent_1995, zanon-willette_ultrahigh-resolution_2011}. Other effects, such as Rabi pulling~\cite{micalizio_Raman-Ramsey_2019} or off-resonant light shifts~\cite{pollock_ac_2018} may contribute to the tuning with $\omegaratio$.

The magnitude of $\chi$ and its dependence on the laser power predicted for a single atomic detuning in Fig~\ref{fig:theory}(b) do not quantitatively match the values measured in the experiment. The measured fluorescence in our CSABC arises from atoms with a distribution of transverse velocities set by our atomic beam divergence and the imaging geometry, and knowledge of the distribution of atomic detunings is needed to make quantitative comparison to the theory~\cite{pati_computational_2015}.  We expect this detuning sampling to reduce the observed magnitude of $\chi$ and the effective values for $\Omega^{2}\!S$. To approximately model this effect, we first calculate the relative clock frequency pulling weight $W$ for atoms with different velocity magnitudes $v$ using the atomic flux distribution $F \propto v^3 e^{-v^2/ \sigma_{v}^2}$ and the Ramsey time $T_{\rm R} \propto 1/v$, where $\sigma_v^2 = 2k_B T/m $, $m$ is the mass of $^{87}$Rb, and $k_B$ is Boltzmann's constant. Considering the $\Delta x =  1 \ {\rm mm}$ region over which we collect fluorescence and the position-velocity correlation observed in our beam, $W(\delta) \sim  e^{-\delta^2/\sigma_{\delta}^2}$, where $\sigma_{\delta } = k\sigma_v \Delta x / 2L \approx 3 \Gamma$ as shown in the shaded are in Fig.~\ref{fig:theory}(a).

Using $W(\delta)$, we estimate the average laser detuning sensitivity of our CPT clock across the measured atoms as  
\begin{equation}
\lrangle{\chi} = \frac{\partial }{\partial \delta} \lrangle{ \phi_{\rm res}/T },
\label{eq:<chi>}
\end{equation}
where $\lrangle{}$ indicates an average weighted by $W(\delta)$. The resulting $\lrangle{\chi}$ is shown in Fig.~\ref{fig:theory}(c) and shows that both $\rho$- and $r$-based resonant shifts persist to higher driving powers than the single-detuning $\chi$ modeling suggests, as the sensitivity is dominated by partially-pumped atoms at finite detuning. For the $r$-based resonant shifts, the magnitude of $\lrangle{\chi}$ and the power at which this sensitivity peaks roughly agree with the data in Fig.~\ref{fig:power}, supporting our claim that resonant light shifts are largely responsible for the observed suppression of $\chi$ in this system. A more quantitatively complete model for $\lrangle{\chi}$ would likely include the multi-level dynamics that modify $r$ as well as the details of the beam geometry and the signal readout accounting for the spatial extent of the atom beam micro-capillary collimator array. 

\section{Doppler Shift Mitigation}

\begin{figure}[ht]
\centering
\includegraphics[width=\linewidth]{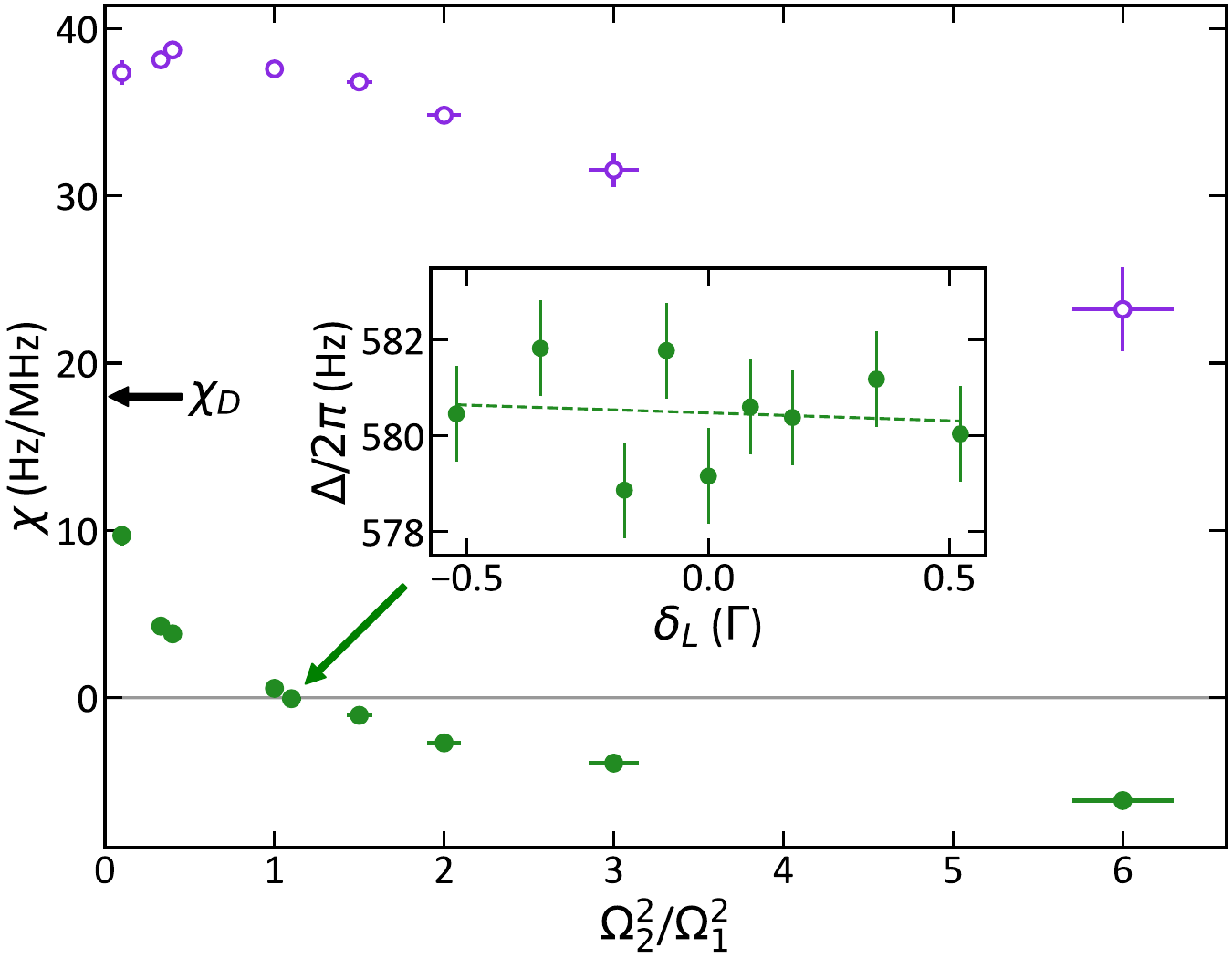}
\caption{ Laser detuning sensitivity $\chi$ vs. $\omegaratio$ for  $\fprime = 1$ (open circles) and $\fprime = 2$ (solid circles) with $P_{\rm res} \approx  113 \ \mu{\rm W}$. Inset shows the nulled clock frequency dependence on common-mode laser detuning near $\omegaratio = 1.1$.}
\label{fig:chi_v_beta}
\end{figure}

We demonstrate tuning of the clock's laser frequency sensitivity by varying $\omegaratio$ at fixed $P_{\rm res}$ as shown in Fig.~\ref{fig:chi_v_beta}. The behavior of the $\fprime = 1,2$ excited states again differs greatly, with $\chi_{\rm res}$ adding or subtracting from the Doppler shift. We observe that $\chi$ generally decreases with increasing $\omegaratio$, which we attribute to modifications of $r$ set by the altered steady-state occupations of the ground state levels. To explain this trend, we note that increasing $\omegaratio$ should drive population into the F = 1 level, which implies an increasing value for the effective value of $r$ and a decreasing value for $\chi_{\rm res}$, consistent with the observed data.

The laser detuning sensitivity passes through zero for $\fprime = 2$, and we tune $\omegaratio$ to find this operational point with $P_{\rm res} \approx 113 \ \mu{\rm W}$. The best value is found near $\omegaratio = 1.1$, where $\chi = (-0.06 \pm 0.21) \ {\rm Hz/MHz}$ as shown in the inset of Fig.~\ref{fig:chi_v_beta}. This value is competitive with other values of $\chi$ reported in the literature \cite{hemmer_ac_1989, blanshan_light_2015} and represents a $> 100$ times rejection of the laser frequency sensitivity relative to the naive $\chi_{\rm D}$ based solely on Doppler shifts. It is convenient that this crossing happens for $\fprime = 2$, as the transition oscillator strength is 3 times higher using this excited state relative to $\fprime = 1$ and the observed CPT contrast is higher for equivalent experimental parameters~\cite{vanier_atomic_2005}. This operation point could likely be adopted in CPT clocks operating using half-hyperfine modulation where $\omegaratio \approx 1$.

\begin{figure}[ht]
\centering
\includegraphics[width=\linewidth]{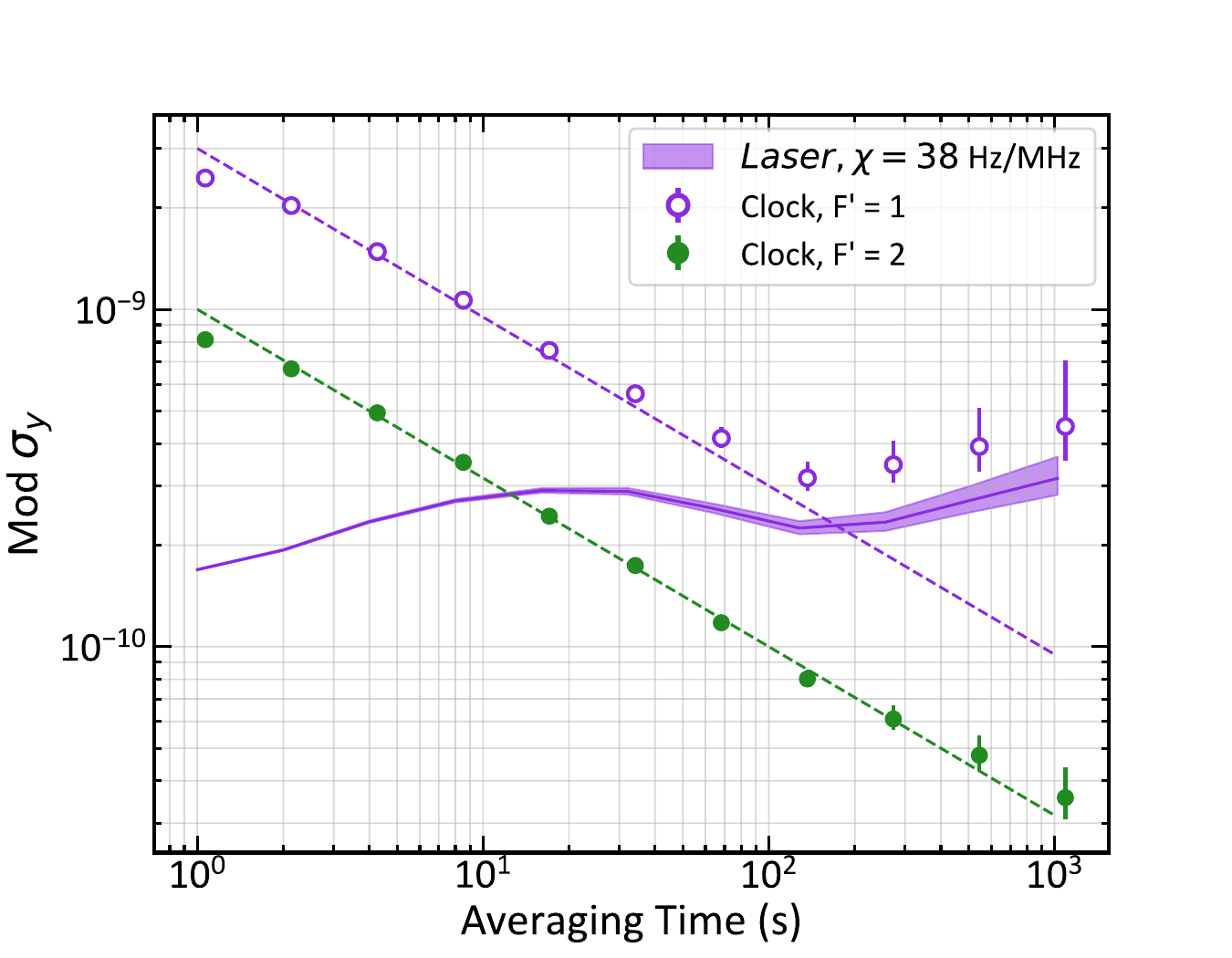}
\caption{Clock stability comparison at optimal operation point for $F^{\prime} = 2$. The clock stability (modified Allan deviation, ${\rm Mod} \ \sigma_{y}$) is measured vs. averaging time using  $\Omega_2^2/\Omega_1^2 = 1.1$ and $P_{\rm res} = 113 \ \mu{\rm W}$ to compare $F^{\prime}  = 2 $ (solid circles, $\chi \approx 0$) and $F' = 1$ (open circles, $\chi \approx 38 \ {\rm Hz/MHz}$). The $F^{\prime} = 1$ laser-detuning-limited stability (shaded band) is plotted using the measured detuning sensitivity and laser frequency stability, explaining the drift of this data set. The $F^{\prime} = 2$ data shows better short-term stability and stable averaging over 1000 s.} 
\label{fig:adev}
\end{figure}

To show the benefit of this operational point, we measure the clock stability using both excited states near $\omegaratio = 1.1$ as shown in Fig.~\ref{fig:adev}. At this operational point, the $\fprime$ = 2 Ramsey fringe (see Fig.~\ref{device_schematic}(d)) is $\approx 30 \ {\rm pW}$ in amplitude and the clock has a fractional frequency instability at 1 s (modified Allan deviation) ${\rm mod}\  \sigma_y \approx 10^{-9}$. The clock stability is measured by recording the clock-stabilized EOM modulation frequency values referenced to a local maser. The clock frequency integrates as $\tau^{-1/2}$ for 1000 s, and is not limited by laser frequency instability over this integration period. For $\fprime$ = 1, the Ramsey fringe signal is $\approx 10 \ {\rm pW}$ and the short-term stability is $\approx 3$ times worse due to the lower overall signal contrast.  The long-term stability of the $\fprime = 1$ lock is degraded by the laser frequency drift beyond 100 s, due to its higher sensitivity to laser detuning.

The laser light is stabilized to a conventional, evacuated vapor cell using modulation-transfer spectroscopy, separate from the spectroscopy on the atomic beam. We characterize the laser frequency stability by recording the optical beat note with a cavity-stabilized optical frequency comb. The measured laser frequency stability is $\approx 55 \ {\rm kHz}$ over integration times from 1 s to 1000 s, placing limits on the clock stability at the $3\times10^{-10}$ level for $\fprime = 1$ ($\chi \approx 38 \ \rm{Hz/MHz}$) and at the $2\times10^{-12}$ level for $\fprime = 2$ ($\chi < 0.2 \ \rm{Hz/MHz}$) over this integration range. We infer the laser-frequency limited stability for $F^{\prime} = 1$ (shaded band in Fig.~\ref{fig:adev}) using the measured sensitivity $\chi = 38 \ \rm{Hz/MHz}$ and the measured laser frequency stability. The agreement between the $\fprime = 1$ clock instability and the limitation set by the laser frequency instability beyond 100 s indicates that laser frequency fluctuations are responsible for the observed degradation in clock performance.


The stability for $\fprime = 2$ the clock integrates to  $\sigma_y \approx 3.5 \times 10^{-11}$ at 1000~s, beyond which the clock stability appears limited by thermal fluctuations of the environment surrounding the clock. The $F^{\prime} = 2 $ performance  is $\approx 10$ times improved relative to $\fprime = 1$ at 1000 s (see Fig.~\ref{fig:adev}), and suppressing the laser frequency sensitivity to the low $10^{-12}$ level indicates that $\mu {\rm s}$-level timing holdover beyond a day is possible using this approach. At elevated clock temperature, near $390 \ K$, the signal height and signal contrast are improved with clock stability mod $\sigma_y \approx  6\times10^{-10}$ at 1 s. Operating at a nominal $\chi \approx 0$ point, using a different $P_{\rm res}$ and $\omegaratio \approx 3$, the clock integrates to the low $10^{-11}$ level with no indication of laser-frequency-induced drift. 

\section{\label{conclusions&discussion} Discussion}
Utilizing the resonant light shifts induced by CPT-Ramsey spectroscopy in our system, we find an approach that can be used to null the clock sensitivity to laser frequency fluctuations. These resonant light shifts are found to agree qualitatively with predictions based on imbalanced decay in the CPT $\Lambda$-system, with the magnitude of the effect explained by the distribution of Doppler detunings probed in our cm-scale beam system. We have also demonstrated that coarse control of the laser detuning sensitivity $\chi$ is achieved through proper choice of excited state, while fine control is achieved by tuning the optical pumping rate and Rabi frequency ratio. The existence of two control parameters ensures tunability for the laser frequency sensitivity even in cases when one of the parameters must remain fixed (e.g. half-hyperfine modulation where $\omegaratio \approx 1$). This implies that our Doppler cancellation scheme is relatively flexible and may be useful in a number of clock configurations where suppression of laser frequency sensitivity is desired.

A laser frequency sensitivity of $\chi~<~0.2 \ \rm{Hz/MHz}$  is easily achieved in multiple experimental configurations with $\fprime = 2$. This implies a required laser frequency stability of $\approx 100 \ \rm{kHz}$ to reach the low $10^{-12}$ level for the clock's fractional frequency stability. To maintain $\chi~<~0.2 \ \rm{Hz/MHz}$, we estimate that the laser power and Rabi-rate ratio must be controlled to $\approx 5 \ \%$, depending on the exact combination of parameters chosen to null the laser frequency sensitivity. The ultimate control requirements for laser power and Rabi-rate ratio will likely depend on the off-resonant light shifts in the system, which are independent of laser frequency for reasonable $\delta_L$~\cite{zhu_theoretical_2000, levi_light_2000, pollock_ac_2018}. Further study of these off-resonant shifts is necessary to determine the performance limitations of CSABCs.

More complete modeling is needed to fully characterize the resonant light shifts in this system. Multi-level simulations have been explored extensively in vapor cell-based clocks, and recently, state-multipole measurements have enabled monitoring of the complex ground state population distribution in a CPT system ~\cite{householder_measuring_2025}. Similar results to the current investigation were reported in ~\cite{tsygankov_branching_2024}, where unequal ground state populations were found to induce nonlinear shifts of the clock frequency vs. optical power. Because many light shift suppression techniques rely on a linear dependence of the error signal on the CPT intensity, these nonlinear shifts present a challenge for long-term clock stability.

The cancellation of the transverse Doppler shift is critical for long-term clock stability in our chip-scale geometry. Other methods exist for minimizing the Doppler shift, but they present challenges for use in low size-weight-and-power clocks. The most direct alternative to Doppler mitigation is to perform spectroscopy using counter-propagating fields~\cite{jau_push-pull_2004,donley_cancellation_2013,esnault_cold-atom_2013,elgin_cold-atom_2019}, but this approach relies on a $\pi$ microwave phase shift for the retro-reflected beam, equivalent to a $\approx 2.2 \ {\rm cm} $ path length for $^{87}$Rb. Further collimation of the atomic beam using cascaded collimators~\cite{li_cascaded_2019} could also reduce $\chi_{\rm D}$, but at the expense of decreased atomic fluorescence, which would then require increased device temperature to reach similar short-term clock stability. Additionally, we expect laser linewidth to play a significant role in determining $\chi$ by altering the optical pumping efficiencies for the transverse velocity distribution being interrogated. The results presented here were obtained using a distributed feedback laser whose linewidth is on the order of $\Gamma$, but future work is needed to fully characterize the interplay between the atomic velocity distribution and the laser linewidth in cm-scale atomic beams.

The presented work is a laboratory demonstration of the chip-scale CPT beam clock approach, and realizing weeklong clock stability will involve miniaturization of the clock architecture and characterization of other leading systematics. We have taken steps in this direction using the device presented here, and we observe that variations in laser power, Rabi-rate ratios, magnetic fields, and density-dependent shifts can be controlled at or below the $10^{-12}$ level at $10^4$~s of integration. Future work utilizing miniaturized and integrated components~\cite{lutwak_chip-scale_2007,zhang_ulpac_2019} is needed to realize lower power operation and to explore the long-term stability limits of the approach.
 



\section{Acknowledgment}
This work was supported by NIST and the Defense Advanced Research Projects Agency (DARPA) H6 program. AS and GM acknowledge financial assistance award 70NANB18H006 from the U.S. Department of Commerce, NIST. We thank Susan Schima for assistance with device fabrication. We thank Judith Olson and Mario Gonzalez Maldonado for helpful suggestions on this manuscript. The views, opinions and/or findings expressed are those of the authors and should not be interpreted as representing the official views or policies of DARPA or the U.S. Government. Distribution Statement ``A" (Approved for Public Release, Distribution Unlimited)

\bibliography{refs_DopplerMitigationPaper2025,manual}

\end{document}